\pdfoutput=1

\documentclass[12pt]{article}
\usepackage[
top = 2.5cm, 
bottom = 2.5cm,  
left = 2.65cm,  
right = 2.65cm]{geometry}

\usepackage{latexsym} 
\usepackage{verbatim}
\usepackage{tikz}
\usetikzlibrary{matrix}
\usepackage{color}
\usepackage{graphicx,amssymb,amsfonts,amsmath,amssymb,amscd,amstext, mathrsfs}
\usepackage{graphicx} 
\usepackage{bbm}
\usepackage{dsfont}
\usepackage{xcolor}
\usepackage[colorlinks=true, citecolor=black, linkcolor=black, allcolors=black]{hyperref}   
\usepackage{amsmath}
\usepackage{empheq}
\usepackage{hyperref}

\newlength\dlf

\def\be{\begin{eqnarray}}
\def\ee{\end{eqnarray}}
\def \bea {\begin{equation}}
\def \eea {\end{equation}}

\def \nn {\nonumber}

\def \ra {\rangle}
\def \rr {\raise.35ex\hbox{\small $\prime$}\kern-.17em{\mbox{\large $\imath$}}}
\def \del {\partial}
\def \dels {\partial\kern-.5em / \kern.5em}
\def \As {{A\kern-.5em / \kern.5em}}
\def \Ds {D\kern-.7em / \kern.5em}

\def \a {\alpha}

\def \k {\kappa}

\def\frac#1#2{{#1\over #2}}

\newcommand{\<}{\langle}
\renewcommand{\>}{\rangle}

\def \iffa {\iffalse} 

\def \ed  {\end{document}}

  \def \a {\alpha}
  
\def\nn{\nonumber}
\def \k {\kappa}

\def \ed  {\end{document}}

\DeclareFontShape{OT1}{cmr}{mx}{n}{<->cmr10}{}
\newcommand{\titlefont}{\fontseries{mx}\selectfont}

\usepackage{cite}

\begin{document}

\begin{titlepage}

\begin{flushright} 
\end{flushright}

\begin{center} 
\vspace{1cm}  

{\fontsize{20pt}{0pt}{\titlefont{Lightcone Commutator 
and 
\\
\vspace{0.4cm} 
 Stress-Tensor Exchange in $d>2$ CFTs 
}}}
\vspace{1.4cm}  

 Kuo-Wei Huang 
\\ 
\vspace{0.8cm} 
{\it
Department of Physics, Boston University, \\
Commonwealth Avenue, Boston, MA 02215, USA
}\\
\end{center}
\vspace{2.2cm} 

{\noindent
Motivated by developing a field-theoretic algebraic approach to 
the universal part of the stress-tensor sector 
of a scalar four-point function 
 in a class of higher-dimensional CFTs, we construct a mode operator, ${\cal L}_m$, 
near the lightcone in $d=4$ CFTs and show that it leads to a Virasoro-like commutator, including a regularized central-term.   
 As an example, we describe how to reproduce the $d=4$ 
 single-stress tensor exchange contribution in the lightcone limit by a mode summation.  
A general-$d$ extension is included.  
We comment on  possible  generalizations.
}

\end{titlepage}

\addtolength{\parskip}{0.8ex}
\jot=2 ex

\subsection*{1. Introduction}

The existence of the Virasoro symmetry lies at the heart of two-dimensional conformal field theories.  
Such an infinite-dimensional stress-tensor algebra dictates universal behaviors 
of $d=2$ systems, allowing  computable multi-point correlation functions and critical exponents \cite{BELAVIN1984333}.   
With the developments of the AdS/CFT correspondence \cite{Maldacena:1997re, Witten:1998qj, Gubser:1998bc} and the revitalized 
conformal bootstrap program (for a review and references see, e.g., \cite{Poland:2018epd}), 
  there has been significant recent progress in using $d=2$ Virasoro conformal blocks to understand quantum 
entanglement \cite{Hartman:2013mia, Asplund:2014coa, Headrick:2015gba},  
chaotic dynamics \cite{Roberts:2014ifa, Fitzpatrick:2016thx, Anous:2019yku}, 
the eigenstate thermalization hypothesis \cite{Lashkari:2016vgj, Faulkner:2017hll, Besken:2019bsu}, 
and to describe gravitational effects in AdS$_3$ with a black hole \cite{Fitzpatrick:2014vua, Fitzpatrick:2015zha, 
Hijano:2015qja, Hijano:2015rla, Anous:2016kss, Fitzpatrick:2016ive, Chen:2017yze, Collier:2019weq}.      

In $d>2$,  the conformal group is finite-dimensional and stress tensors generally do not form an algebra. 
One should not expect to find a model-independent way that universally captures the full stress-tensor contributions. 
Is it therefore completely hopeless if one desires to generalize a similar story to $d>2$ in a certain way? 
 
The motivation of the present work comes from recent 
growing evidence \cite{Fitzpatrick:2019zqz, Huang:2019fog, 
Fitzpatrick:2019efk, Kulaxizi:2019tkd, Li:2019tpf, Karlsson:2019dbd, Li:2019zba, Karlsson:2019txu} 
indicating that a certain universality of multi-stress 
 tensors in a class of $d>2$ CFTs   appears  in the limit where 
operators in a correlator approach each other's lightcone or, equivalently, in the lowest-twist limit.
 We mainly focus on CFTs with a large central charge 
and a four-point function with two heavy and two light scalars.  
 The universality, more precisely, means that the operator product expansion  (OPE)
coefficients of the lowest-twist multi-stress tensors are protected, in the sense 
that they are fixed by dimensions of scalars and the central charge $C_T$. 
 The higher-twist OPE coefficients, on the other hand, can 
be contaminated by more model-dependent parameters. 
From the gravity side's viewpoint, this implies that the lowest-twist OPE coefficients are insensitive to  
higher-curvature terms in the purely gravitational action, i.e. they may be determined by Einstein gravity.   
In $d=2$, such universality can be explained by the Virasoro symmetry. 
The recent $d>2$ results  share intriguing similarities with $d = 2$ CFTs and 
we are motivated to search for a Virasoro-like derivation in 
 the lightcone limit in $d > 2$ CFTs; we largely focus on $d=4$ in this note.

A recent effort toward this direction was made in \cite{Huang:2019fog}. 
Based on the most general stress-tensor commutators consistent 
with the Poincar\'e algebra in local QFT \cite{doi:10.1063/1.1705368}, 
 it was shown that, under an assumption on the Schwinger term, a 
Virasoro-like stress-tensor commutator emerges near the lightcone in $d=4$ CFTs.  
Here, we would like to start to build a bridge between the  stress-tensor commutator 
and the conformal block decomposition of a scalar four-point function.  
We will also remark on the Schwinger-term assumption.   

To build a bridge between the    
stress-tensor commutator and the scalar 
correlator, it is desirable to construct an effective mode operator, similar to the generator $L_m$ in $d=2$.   
An immediate obstacle, however, is that the $d=4$ stress-tensor commutator 
has a UV cutoff, $\Lambda$, dependent central-term. ($\Lambda$ has mass dimension one.)
We will propose an ${\cal L}_m$, defined near the lightcone, and show that, using the   
 stress-tensor commutator, it results in a Virasoro-like $[{\cal L}_m , {\cal L}_n]$.   
The basic picture of this construction is that we  treat the additional two-dimensional transverse space as a thin layer 
with a thickness defined by a short-distance cutoff $\epsilon$. 
 The product $\epsilon^2  \Lambda^2$ gives a dimensionless finite constant.  
Introducing a thin region, instead of infinite transverse space, may be interpreted 
as a part of the lightcone limit, where we arrange scalars to live on a $d=2$ plane and the stress tensors contribute only near the plane. 

We will describe how to use 
 the mode operator ${\cal L}_m$  to compute the single-stress tensor exchange in the lightcone limit by a direct mode summation.  
This computation generalizes the Virasoro-algebra derivation of the one-graviton contribution to the 
identity block in $d=2$ CFTs described in \cite{Fitzpatrick:2014vua}. 
 
The more general case, beyond single-stress tensor, is more involved partially 
because the stress-tensor-scalar, $T{\cal O}$, OPE in higher dimensions has a delicate structure. 
While the general story is left to future work, we will make some preliminary remarks on a possible multi-stress tensor generalization.

\subsection*{2. A lightcone mode operator}

\subsubsection*{{\it 2.1.  Stress-tensor commutator near the lightcone}}

We start with the stress-tensor commutation relation  in $d=4$ CFTs in 
 Minkowski spacetime $ds^2 = -dx^+ dx^- + dy^2+ dz^2$ where $x^\pm= t\pm x$.    
Using the tracelessness condition, one can  write the relevant 
component of the stress tensor in the lightcone limit as 
$T^{++}=-2 (T^0_0-T^0_1)- T^a_a$; $a=2,3$ denote transverse directions.  
An important point is that the purely-spatial components of the stress tensor generally 
do not admit a model-independent commutator \cite{doi:10.1063/1.1705368}.   
However, in the case where stress tensors are inserted 
in a scalar correlator, the transverse components are suppressed in the lightcone limit. 
(By lightcone limit, we mean that we consider 4 scalars to lie on an $x^+ - x^-$ plane with 
 configuration $\<{\cal O}(\infty) {\cal O}(1) {\cal O} (x^+, x^-)  {\cal O} (0) \>$, 
and then take $x^- \to 0$. 
We also send stress tensor's $x^-_{T} \to 0$.)  
The dominating contribution in the lightcone limit is \cite{Huang:2019fog}
\be
\label{4dVira}
  &&-i[\widetilde T^{++}(x^+, x^a), \widetilde T^{++}(x'^+, x'^a)] =
- 4 \Big(\widetilde T^{++}(x^+, x^a)+\widetilde T^{++}(x'^+, x'^a) \Big)\del_+\delta^{3}\nn\\
&&~~~~~~~~~~~~~~~~~~~~~~~~~~~~~~~~~~~~~~~~~~~
+  {C_T\pi^2 \over 60} \Big( \Lambda^2 +  \Delta  \Big) \del_+ \Delta \delta^3\ , 
\ee  
where $\widetilde T^{++}= -2 (T^0_0-T^0_1)$, $\delta^{3} 
 = \delta(x^+ - x'^+) \delta^{2}(x^a-x'^a)$, and $\Delta$ is a Laplacian. 
Note that  the central-term contains a UV cutoff $\Lambda$-dependent piece.  
 We have set $x^- \to 0$ in the above commutator. More formally,  one can write  
$x^- = \epsilon$ and then focus on the leading small-$\epsilon$ contribution.  

The result \eqref{4dVira} is valid only when the Schwinger term 
 in the stress-tensor commutator is a $c$-number. 
That is, the central-term in \eqref{4dVira} is assumed to be 
the same as the expectation value of the stress-tensor commutator.  
{\it A priori}, however, there might be an additional operator Schwinger term. 
It remains an interesting question to ask in what class of   
CFTs the Schwinger term is effectively a $c$-number (in the lightcone limit) as it may 
be related to the validity of the universality.     

In what follows, we shall simply assume that we focus on the class of 
CFTs where the Schwinger term is effectively 
a $c$-number and adopt \eqref{4dVira}.

\subsubsection*{{\it  2.2.  A mode operator and $[{\cal L}_m,{\cal L}_n]$ in $d=4$}}

To develop a Virasoro-like effective representation theory 
for the class of  higher-dimensional CFTs 
whose lowest-twist subsector 
has a universal meaning, one would like to explore 
possible constructions of an effective mode generator, denoted as ${\cal L}_m$, 
which is defined via  
 integrating the coordinates of a stress tensor out.  
Our goal here is to find an ${\cal L}_m$ such that, when combined with 
the stress-tensor commutator near the lightcone, it can lead to a commutator $[{\cal L}_m,{\cal L}_n]$ 
which  (i) satisfies the Jacobi identity and (ii) has a regularized central-term.   

Since the difference between $\widetilde T^{++}$ and $T^{++}$ is suppressed 
in the lightcone limit, as mentioned, we simply adopt $T^{++}$ in the following 
to have simpler expressions.

Let us Wick rotate to a Euclidean plane,   
 $ds_{(E)}^2
= dx^+_{(E)} dx^-_{(E)}+ dy^2+dz^2$, with  complex coordinates 
 $x^\pm_{(E)} \equiv x^1  \pm i x^2$. (The subscript will be dropped.)     
We keep the extra two-dimensional 
transverse directions uncompactified.\footnote{In general, one can consider other geometries such as a torus.}      
Consider the following ansatz:  
\be
{\cal L}^{j,k}_m=  \lim_{x^{-} \to \epsilon} \int \int dy ~dz ~f(y,z; j, k)~\oint {dx^+\over 2 \pi i}~ (x^+)^{m+1}~ T^{++}(x^+, x^-,y, z) \ .
\ee 
  Changing the power of $x^+$ corresponds to shifting $m$; we adopt $m+1$ for later convenience.  
The smear function $f(y,z; j, k)$ generally can depend on new mode 
numbers, $j, k$, associated with transverse coordinates.  
The integrals along the transverse directions are necessary as the stress-tensor 
commutator contains Dirac delta-functions; just sending $y, z$ 
to zero in the stress tensor does not give a sensible commutator.  

The Jacobi identity severely constrains the form of $f(y,z; j, k)$.  
We propose 
\be
\label{4dL}
{\cal L}_m=   
-  {\pi \over 2}  \lim_{L,x^{-} \to \epsilon} ~\int^{L}_{0}  \int^{L}_{0} dy ~dz ~\oint {dx^+\over 2 \pi i}~ (x^+)^{m+1}~ T^{++}(x^+, x^-, y, z) \ ,
\ee   
where a short-distance scale $\epsilon$ is introduced for the transverse directions.   
 The stress-tensor contribution therefore comes only from a very thin region near a $d=2$ 
 plane.\footnote{One may adopt asymmetric limits, $x^{-} \to \epsilon$, $L \to a \epsilon$. 
While intermediate expressions can then depend on $a$ we will find the stress-tensor exchange final result is independent of $a$; we set $a=1$ for simplicity.    
}   

The stress-tensor commutator \eqref{4dVira}  
and \eqref{4dL} give, in the lightcone limit, 
\be
\label{LL1}
&&[{\cal L}_m , {\cal L}_n]= (m-n) {\cal L}_{m+n} \nn\\
&&~~~~~~~ 
+  
  {C_T\pi^3 \over 480} \Lambda^2 \epsilon^{2}~ m (m^2-1) \delta_{m+n,0}
- 
{C_T \pi^3 \over 480} \epsilon^{2} ~ m (m^2-1)(m-2)(m-3) \delta_{m+n,2} 
\ee    
where the Cauchy integral theorem was used.\footnote{The radial ordering is implicit.  
 In the lightcone limit, $\partial_+ T^{++}$ is also suppressed.  The central-term  is finite and $T^{++}$ is independent of $x^+$ in $d=2$ where  
one can derive the Virasoro algebra of the form $[L_m,  L_n]$ starting with the $d=2$ stress-tensor commutator.   
With additional coordinates, we should be concerned about the passage from equal-time commutators to radial quantization in $d>2$. 
 We shall proceed by assuming that the related  subtleties are insignificant in the limit $\epsilon \to 0$.}  
 In this notation, $\delta_{m+n,0}$ has mass dimension $-m-n$, 
 and $\delta_{m+n,2}$ has dimension $-m-n+2$, both with  magnitude unit.   
Keeping an explicit $\epsilon$ for the limit of $x^{-}$ is irrelevant in deriving \eqref{LL1}, but it will be useful when we later consider 
a scalar correlator with a stress tensor inserted.      

We consider that the large UV cutoff term suppresses the last piece of \eqref{LL1}, which causes tension with the Jacobi identity.  
 The product of the UV cutoff $\Lambda$ and the short-distance 
regulator $\epsilon$ is a dimensionless parameter.\footnote{If one first 
redefines $\Lambda \to \widetilde\Lambda$ in \eqref{4dVira} to absorb $C_T$, one  needs to reintroduce $C_T$ via ${\widetilde\Lambda}^2 \epsilon^2 \sim C_T$. This process looks {\it ad hoc} and we do not adopt it here.  
However,  it is interesting to note that a similar identification appears in 
 the soft-theorem related literature: see (147) in \cite{Cheung:2016iub}.  
There, the central charge is related to internal soft exchanges.  
I thank L.~Fitzpatrick for a discussion.}     
We arrive at a commutator near the lightcone: 
\be
\label{LL}
[{\cal L}_m , {\cal L}_n]= (m-n) {\cal L}_{m+n} + \a ~C_T ~  m (m^2-1) \delta_{m+n,0}   \ .
\ee  
We shall further discuss the arbitrariness of $\a$ in the next section.        
Note that, unlike in $d=2$ where the Virasoro algebra represents an exact symmetry, \eqref{LL} is an effective description: 
in \eqref{LL}, we have ignored contributions suppressed in the lightcone limit and assumed a class of  CFTs with a $c$-number Schwinger term.

\subsection*{3. A Virasoro-like derivation of $d=4$ single-stress tensor exchange}

The lightcone algebra \eqref{LL} looks formally the same as the $d=2$ Virasoro algebra. 
We may assume, in a universal class of $d=4$ CFTs, there exists a   
lowest-twist subsector where the associated
intermediate states, $| \a \>_s$ ($s$ denotes a subspace), 
can be effectively organized into a Virasoro-like  representation theory.   
In some sense, the lightcone limit acts like  picking the 
holomorphic sector out and  we do not need to introduce $``{\cal \bar L}_{m}"$.    

Focusing on such a subspace, we may try to follow the terminology 
 of the highest-weight representation in $d=2$ CFTs: 
  $ {L}_{0} | h \>= h | h \>$ and $ { L}_{m} | h \>= 0$ for  $m \geq 1$. 
The modes ${L}_{m}$ with $m<0$ generate descendants.  
The  vacuum $|0 \>$, preserving the maximal number of symmetries, is the associated state of the identity operator that has $h=0$.     
 One important difference, however, is that $T{\cal O}$ 
OPE structure in $d=4$ is more delicate.\footnote{For explicit expressions at the first few   
orders in the OPE, see, for instance, Appendix B of \cite{Belin:2019mnx}.}
Using the $T{\cal O}$ OPE to express ${\cal L}_m$ as a differential 
operator will not be included in the present note. 
Here, we focus on a Virasoro-like derivation of the {\it single}-stress tensor exchange in the lightcone limit.   
This derivation does not require knowing ${\cal L}_m$ as a differential operator 
because we can use the three-point function $\< T {\cal O} {\cal O} \>$, together with \eqref{4dL}.   

 The three-point function of a stress tensor with two scalar primaries in $d=4$ is \cite{Osborn:1993cr}
\be
\label{TOO}
\< T^{\mu\nu}(x_1) {\cal O}_{\Delta} (x_2) {\cal O}_{\Delta}(x_3) \> 
= {c_{T{\cal O}{\cal O}}\over x^4_{12} x^4_{13} x^{2\Delta-4}_{23}} \Big({X^\mu X^\nu \over X^2} -{\delta^{\mu\nu}\over 4} \Big) \ , ~~~
X^\mu = { x^\mu_{12}\over x^2_{12} }-{ x^\mu_{13}\over x^2_{13} } 
\ee  
with $c_{T{\cal O}{\cal O}}= - {2 \Delta\over 3 \pi^2}$.  
We shall focus on the identity block 
 at large $C_T$ with the heavy-light limit:  $\Delta_H \sim C_T$,  $\Delta_L \sim {\cal O} (1)$.    
 The single-stress tensor exchange contribution, discussed below, may be computed without explicitly imposing these limits, but we will still formally adopt $\Delta_H$ 
and $\Delta_L$ in what follows, having in mind a potential generalization involving multi-stress tensors. 

In the lightcone limit, we assume that the corresponding intermediate states can be  effectively 
generated by the operator ${\cal L}_{m}$ acting on the vacuum.   
Introduce a basis
\be
|\a_0 \>_T={ {\cal L^\dagger}_{m} | 0 \>\over \sqrt{{\cal N}_m}  } \ , ~~~~~~ {\cal N}_m= \< {\cal L}_{m} {\cal L}^\dagger_{m}\> , 
\ee 
which formally represents a normalized one-graviton state. Assume $m>1$ here. 
We may relate the single-stress tensor conformal block in the lightcone limit to the following object: 
\be
\label{V1}
{\cal V}_T = \lim_{\bar z  \to 0} \sum^\infty_{m=2} {\<{\cal O}_H(\infty) {\cal O}_H(1) {\cal L}^\dagger_{m} \>\<{\cal L}_{m}{\cal O}_L (z, \bar z ){\cal O}_L(0)  \> \over {\cal N}_m  
 ~\<{\cal O}_H(\infty) {\cal O}_H(1)  \>\<{\cal O}_L (z, \bar z ){\cal O}_L(0)  \> } \ .
\ee 
We have adopted the conventional variables $z, \bar z$ defined 
by $u= z \bar z$, $v= (1-z)(1-\bar z)$ where $u, v$ are conformal cross-ratios.  
 In the configuration \eqref{V1}, $x^+= z$, $x^-= \bar z$ for the scalar and the scalar's lightcone limit is $\bar z \to 0$.   
The Hermitian conjugate of the $d=4$ stress tensor in the radial quantization is 
\be
\label{dagger}
T^{\mu\nu} (x)^\dagger 
=   {\cal I}^\mu_{~\lambda}(x) {\cal I}^\nu_{~\rho}(x)  x^{-8} T^{\lambda\rho} \big({x\over x^2}\big) \ , 
~~~~{\cal I}^\mu_{~\lambda}(x)= \delta^\mu_\lambda- 2 { x^\mu x_\lambda \over x^2} \ .
\ee 
Let us first compute the numerator of \eqref{V1}  using \eqref{dagger}, \eqref{TOO}, and \eqref{4dL}. The computation is short but can be thorny as it involves a certain order of limits. 

Denote $y_T, z_T$ as the transverse coordinates for $T^{++}$ and $r^2=y_T^2+z_T^2$.  We have 
\begin{align}
{\<{\cal L}_{m}{\cal O}_L (z, \bar z ){\cal O}_L(0)  \> \over  \< {\cal O}_L (z, \bar z ){\cal O}_L(0) \>}
&=
- {\pi\over 2} \lim_{L, x^- \to \epsilon} ~ \int^{L}_{0}  \int^{L}_{0} dy_T ~dz_T ~{\< \oint {dx^+\over 2\pi i} 
~ (x^+)^{m+1}~ T^{++}(x^+, x^-, y_T, z_T) {\cal O}_L (z, \bar z ){\cal O}_L(0)  \> \over  \< {\cal O}_L (z, \bar z ){\cal O}_L(0) \>}\nn\\
&= 
\Delta_L ~
 \lim_{L, x^- \to \epsilon} ~  \int^{L}_{0}  \int^{L}_{0} dy_T ~dz_T  ~~ 
\Big [ 
r^4   
\big( z - {r^2\over x^-} \big)^{m-1} \\
&~~~~~\times
{ (m-2)(m-3) (x^-)^2 z^2 
+ 6 (m-3) r^2 x^- z 
 + 12 r^4 
\over 6 \pi ~(x^-)^8 z^2  } \Big ]\bar z + {\cal O}(\bar z^2) \ , \nn
\end{align}
in a small $\bar z$ expansion.  
In performing the contour integral, we have picked the pole due to ${\cal O}_L (z, \bar z )$.    
We should consider 
 the stress tensor's lightcone limit as $x^-\to \epsilon$, instead of directly setting  $x^-=0$ from the start
 in this correlator computation.   
After performing the remaining integrals,
\begin{align}
\label{A}
{\<{\cal L}_{m}{\cal O}_L (z, \bar z ){\cal O}_L(0)  \> \over  \< {\cal O}_L (z, \bar z ){\cal O}_L(0) \>}
&=     { 14  \Delta_L (m-2)(m-3) \over 135 \pi  }  z^{m-1}   \bar z   ~+~ \rm {subleading} \ .
\end{align}
A similar procedure, using \eqref{dagger}, gives\footnote{Here $T^{--}$ contributes instead due to the projector in \eqref{dagger}.  The corresponding lightcone limit becomes  $x^+ \to \epsilon$ while $x^-$ is integrated out.  Note the stress-tensor part  involves an inversion, $x \to {x\over x^2}$, implying $x^+ \to {1\over \epsilon}$ for the stress-tensor part.  We again focus on the contribution near a $d=2$ plane.}  
\be
\label{B} 
{\<{\cal O}_H(\infty) {\cal O}_H(1) {\cal L}^{\dagger}_{m} \> \over  \< {\cal O}_H (\infty){\cal O}_H(1) \>} 
=   { \Delta_H (m-2)(m-3) \over 6 \pi  } ~+~ \rm {subleading} \ .
\ee   
As the leading contributions vanish at $m=2, 3$,  
we may interpret that the vacuum is annihilated effectively by the operators ${\cal L}^\dagger_{2}$ and  ${\cal L}^\dagger_{3}$.  
By effective, we mean ${\cal L}_m$ is in a correlator with the lightcone limit imposed. 
In $d=2$, the corresponding \eqref{A} and \eqref{B} both give a factor of $(m-1)$, but the lightcone limit is not necessary. 

The computation of  the normalization factor  is more involved than the $d=2$ case where  $L^\dagger_{m}= L_{-m}$.  
While we do not find the same relation for the lightcone mode operator ${\cal L}_m$ in $d=4$, we can still compute ${\cal N}_m$ via the stress-tensor two-point function: 
\begin{align}
\label{TT4d}
\langle T^{\mu\nu} (x_1)T^{\lambda\rho} (x_2)\ra
&= C_T  {{\cal I}^{\mu\nu,\lambda \rho}(s)\over s^{8}}\ , ~~~ s= x_1- x_2  \ ,  \nn\\
{\cal I}^{\mu\nu,\lambda \rho}(s)&= {1\over 2} \Big( {\cal I}^{\mu\lambda}(s) {\cal I}^{\nu\rho}(s)+ {\cal I}^{\mu\rho}(s) {\cal I}^{\nu\lambda}(s)\Big)-{1\over 4} \delta^{\mu\nu} \delta^{\lambda \rho}    \ .
\end{align}
A subtlety appears when imposing the lightcone limit on both stress tensors. 
More generally, one can consider cutoffs $\epsilon_1 \equiv \epsilon$, $\epsilon_2 \equiv \gamma \epsilon$, for the first and second stress tensors, respectively. 
 Using \eqref{TT4d}, \eqref{dagger}, and \eqref{4dL},  we find
\be
\< {\cal L}_{m} {\cal L}^\dagger_{m}\>= \< [{\cal L}_{m}, {\cal L}^\dagger_{m}]\>= {7 \pi^2 \over 1350 (1-\gamma)^6} ~C_T~ m \big(m^2-1\big)\big(m-2\big)\big(m-3\big) ~+~ \rm {subleading} \ .
\ee  
In general, the slope $\gamma$ should be allowed to be zero or an arbitrary nonzero constant so that two cutoffs, 
$\epsilon_1$ and $\epsilon_2$, are independent of each other. 
However, here we have a pole at $\gamma=1$, which reflects the divergence appearing when the two stress tensors are on exactly the same lightcone. 
This implies that a given nonzero $\epsilon_2$ would forbid $\epsilon_1$ to be the same value and, in this sense, they are correlated. 
The finite result and noncorrelated cutoffs requirements thus fix $\gamma=0$. 
Note we set $\gamma=0$ at the end of the computation. 
 
  We obtain, in the lightcone limit, ($\k(\gamma)= { 10 (1-\gamma)^6 \over 3 \pi^4}  $)   
\begin{align}
\label{4dsum}
{\cal V}_T
&=  \k(\gamma)  ~{\Delta_H \Delta_L \over C_T}\sum^\infty_{m=4} { (m-2)(m-3) \over m(m^2-1)}~ z^{m-1}  \bar z \\
&= \k(\gamma)  ~{\Delta_H \Delta_L \over C_T}{3 (z-2) z- \big(6 +(z-6)z \big) \ln(1-z) \over z^2} \bar z 
=  {\k(\gamma)\over 30} ~{\Delta_H \Delta_L \over C_T} z^3 ~ {}_2 F_1(3,3,6,z)  \bar z \ , \nn
\end{align}
which is the $d=4$ single-stress tensor block in the lightcone limit.  
We shall require $\gamma=0$, as mentioned, and this also gives the conventional single-stress tensor OPE coefficient \cite{Dolan:2000ut}.\footnote{See also \cite{Karlsson:2019dbd} for related conventions. Note $C_{T {\rm (there)}}= \Omega_{d-1}^2  C_{T {\rm (here)}}$ with $\Omega_{d-1}= {2 \pi^{d\over 2}\over  \Gamma({d\over 2})}.$}

We emphasize that the overall coefficient of \eqref{4dsum} is insensitive to additional rescalings related to transverse directions: if one replaces $\int^{L}_{0}  \int^{L}_{0} dy dz$ in \eqref{4dL} with $\int^{a L}_{0}  \int^{b L}_{0} dy dz$, the $a,b$ dependences cancel out in the final correlator; we set $a=b=1$ to have simpler intermediate expressions. 
On the other hand, the central-term of $[{\cal L}_m, {\cal L}_n]$ can depend on $a, b$ and thus the value of $\alpha$ in \eqref{LL} remains arbitrary.  
However, although a Virasoro-like effective algebra provides a justification for the near-lightcone operator ${\cal L}_m$, the commutator $[{\cal L}_m, {\cal L}_n]$ is not explicitly used in the above particular computation.  Indeed, we have shown that the operator ${\cal L}_m$, the three-point function \eqref{TOO}, and the stress-tensor two-point function are sufficient to compute the stress-tensor exchange.   A direct generalization, discussed below, will also allow us to capture the general $d$ stress-tensor exchange. 

\subsection*{4. General $d$ stress-tensor exchange from  ${\cal L}_{m}$} 

We define
\be
\label{alldL}
{\cal L}_m=   
-  {\pi \over 2}  \lim_{L,x^{-} \to \epsilon} \int^{L}_{0}  \cdots \int^{L}_{0} dx^{\perp}_1  ~\cdots~dx^{\perp}_{d-2} ~\oint {dx^+\over 2 \pi i}~ (x^+)^{m+1}~ T^{++}(x^+, x^-, x^{\perp}_A) \ ,
\ee   where $A=(1,2,..., d-2)$.   This is a straightforward extension of the $d=4$ case.    

The structure is simpler in even dimensions where one can avoid fractional exponents and thus we restrict to even $d$ to search for a general pattern.
We find the following leading contributions:
\be
\lim_{\bar z\to 0}  {\<{\cal L}_{m}{\cal O}_L (z, \bar z ){\cal O}_L(0)  \> \over  \< {\cal O}_L (z, \bar z ){\cal O}_L(0) \>}   &\sim&  \Delta_L  {\Gamma(m+1-{d\over 2})\over \Gamma(m+1-d)}~ z^{m-{d-2\over 2}}  \bar z^{d-2\over 2}\ , \\
 {\<{\cal O}_H(\infty) {\cal O}_H(1) {\cal L}^{\dagger}_{m} \> \over  \< {\cal O}_H (\infty){\cal O}_H(1) \>} &\sim&  \Delta_H  {\Gamma(m+1-{d\over 2})\over \Gamma(m+1-d)} \ , \\
 \< {\cal L}_{m} {\cal L}^\dagger_{m}\> &\sim&  C_T  {\Gamma(m+2)\over \Gamma(m+1-d)} \ .
\ee 
The summation reproduces the near-lightcone single-stress tensor exchange structure:
\be
 \sum^\infty_{m=d} {\Gamma^2(m+1-{d\over 2})~ z^{m-{d-2\over 2}}  \over   \Gamma(m+2) \Gamma(m+1-d) } \bar z^{d-2\over 2}={{\sqrt\pi} \Gamma({d+2\over 2})\over 2^{d+1} \Gamma({d+3\over 2})} ~z^{d+2\over 2}~ {}_2 F_1\big({d+2\over 2},{d+2\over 2}, d+2, z\big)~ \bar z^{d-2\over 2} \ .
\ee  
To keep the expressions simple, we only emphasized the $m$-dependence.  
As before, $ \< {\cal L}_{m} {\cal L}^\dagger_{m}\>$ depends on a parameter $\gamma_d$.   
We have verified that, like the $d=4$ case, setting $\gamma_d=0$ reproduces the conventional single-stress tensor OPE coefficient.

\subsection*{5. Concluding remarks and outlook}

We have described a derivation of the near-lightcone single-stress tensor 
 block in $d>2$ CFTs via a Virasoro-like generator. 
The lightcone mode operator ${\cal L}_m$ is defined by integrating the $d>2$ stress tensor near a $d=2$ plane where scalars live. 
This picture also suggests a way to deal with the UV divergence in $d>2$ stress-tensor commutators.

In a recent wok \cite{Karlsson:2019dbd}, an ansatz has been proposed 
for the multi-stress tensor sector of the heavy-light scalar correlator in the lightcone limit. 
Assuming such an ansatz, the resulting OPE coefficients agree 
 with the earlier holographic computation \cite{Fitzpatrick:2019zqz}.  
The proposed near-lightcone ansatz can be expressed as a sum 
of products of hypergeometric functions, which is quite similar to the $d=2$ Virasoro vacuum block.  
It would be interesting to derive such a pattern involving 
multi-stress tensors in $d>2$ CFTs based on a Virasoro-like approach.   
Being optimistic, the fact that we are able to reproduce the $d>2$  
single-stress tensor block near the lightcone via ${\cal L}_m$  
 perhaps hints that an algebraic derivation for the more general case exists.  
It would be interesting to further develop an effective representation theory near the lightcone for this universal class of $d>2$ CFTs, and also search for possible extensions.

A potentially important step, which we have not considered in the present note,  
is to express ${\cal L}_m$ as a differential operator.    
In \cite{Belin:2019mnx}, using the $T {\cal O}$ OPE, 
the authors show how to recast the $d=4$ averaged null energy (ANEC) operator as a 
differential operator, given as a series expansion and then resum.   
(See also \cite{Casini:2017roe, Cordova:2018ygx, Kologlu:2019mfz, Rosso:2019txh, Balakrishnan:2019gxl} for related discussions.)  
Note that the ANEC operator can be related to ${\cal L}_{-1}$ after integrating coordinates out. 
Considering a more general case to obtain a differential-operator form of 
${\cal L}_m$ in $d>2$ can be useful.    

A differential-operator form of ${\cal L}_m$ should 
in principle allow one to compute the following more general object:
\be
\label{Vk}
{\cal V}_{T^k} = \lim_{\bar z  \to 0} {\<{\cal O}_H(\infty) {\cal O}_H(1) ~{{{\cal P}^{(k)}}}
 ~ {\cal O}_L (z, \bar z ){\cal O}_L(0)  \> \over ~\<{\cal O}_H(\infty) {\cal O}_H(1)  \>\<{\cal O}_L (z, \bar z ){\cal O}_L(0)  \> } \ ,
\ee 
with the $k$-stress tensors  lightcone  projector ${\cal P}$ given by 
\be
{{{\cal P}^{(k)}} } 
= \sum_{\{m_i, k_i\}} \frac{{\cal L}_{m_1}^{k_1 \dagger} 
\cdots {\cal L}_{m_n}^{k_n \dagger}  |0\>\<0| {\cal L}_{m_n}^{k_n} 
\cdots {\cal L}_{m_1}^{k_1}}{{\cal N}_{\{m_i, k_i\}}} \ .
\ee  
A direct $k>1$ computation is more complicated, but one may expect that, similar to the $d=2$ case, the computation 
can be simplified in the geodesic limit, $\Delta_L \to \infty$, leading to a possible exponentiation in the lightcone limit.  
Moreover, it might be possible to derive certain  near-lightcone  null-state equations in a class of $d>2$ CFTs via an algebraic approach. 
 We hope to discuss these possibilities somewhere else.

Let us end by mentioning another general question that has not been addressed: 
what is the validity of the lightcone universality in  $d>2$ CFTs?\footnote{We expect that the correlator generally becomes less sensitive to model-dependent details in the lightcone limit; in the simplest class of CFTs, the lowest-twist OPE coefficients depend only on $\Delta_H$, $\Delta_L$, and $C_T$.  It would be interesting to incorporate possible extra parameters for a wider class of $d>2$ CFTs.}  

\vspace{0.2 cm} 

\begin{center} 
\subsubsection*{Acknowledgments}
\end{center} 

{\noindent 
I would like to thank H.~Casini, L.~Fitzpatrick, T.~Hartman, C.~Herzog, and A.~Tseytlin for helpful comments.
This work was supported in part by the Simons Collaboration grant on the Non-Perturbative Bootstrap and  
in part by the U.S. Department of Energy Office of Science  under award No.  
DE-SC0015845. Opinions and conclusions expressed in this work are those of the author and do not necessarily reflect the views of funding agencies.
}

\vspace{1 cm} 

\bibliographystyle{utphys}
\bibliography{4dLCBib}

\end{document}